# Light-emission from ion-implanted group-IV nanostructures

Moritz Brehm

*Institute of Semiconductor and Solid State Physics, Johannes Kepler University Linz, Altenberger Strasse 69, 4040 Linz, Austria*

Email: moritz.brehm@jku.at

**Abstract**

Silicon photonics is destined to revolutionize technological areas, such as short-distance data transfer and sensing applications by combining the benefits of integrated optics with the assertiveness of silicon-based microelectronics. However, the lack of practical and low-cost silicon-based monolithic light sources such as light-emitting diodes and, in particular, lasers is the main bottleneck for silicon photonics to become the key technology of the 21$^{st}$ century. After briefly reviewing the state of the art regarding silicon-based light-emitters, we discuss the challenges and benefits of a highly flexible approach: The epitaxial incorporation of group-IV nanostructures into crystalline silicon. We argue that a paradigm change for group-IV quantum dots (QDs) can be achieved by the intentional incorporation of extended point defects inside the QDs upon low energy ion implantation. The superior light-emission properties from such defect-enhanced quantum dots (DEQDs), our present understanding of their structural formation and light-emission mechanisms will be discussed. We will show that useful electrically-driven devices, such as light-emitting diodes (LEDs) can be fabricated employing optically active DEQD-material. These LEDs exhibit exceptional temperature-stability of their light emission properties even up to 100°C, unprecedented for purely group-IV-based optoelectronic devices. Thereafter, we will assess the superior temperature stability of the structural properties of DEQDs upon thermal annealing, the scalability of the light-emission with the DEQD density and passivation schemes to further improve the optical properties. The chapter ends with a discussion of future research directions that will spark the development of this exciting field even further.

## 1. Introduction to the chapter

Silicon (Si), the dominating material for microelectronics [1], is generally a poor light-emitter which is mainly owed to the indirect nature of its energy bandgap. Nevertheless, silicon-based photonics (Si photonics) is rising to be a key technology for the 21$^{st}$ century and the implementation of practical and low-cost light sources based on group-IV nanostructures [2-9] and group-IV alloys [10-13] can ultimately be significant to this paradigm change. Si Photonics is expected to solve challenges, such as *e.g.* low-cost, high-bandwidth and CMOS-compatible optical interconnects with low energy consumption [14,15] or driving applications for sensing and lab-on-a-chip-technologies [16,17] with access to the manufacturing power of CMOS integrated circuits.

Given the fundamental limitations of Si as a light source, a wealth of strategies has been explored to develop material combinations capable of enhancing light emission (see further discussion in sections 1.3 and 1.4). Among others, the introduction of germanium (Ge) nanostructures in Si (Ge/Si) sparked hopes and intensive research in various directions for the last three decades [2-9]. While self-assembled epitaxial Ge quantum dots (QDs) have been successfully used to study the rich physical phenomena leading to their nucleation [2,6,7,9,18], their performance as light-emitters has remained disappointingly low due to three key factors:

i) Both, Si and Ge are indirect semiconductors.
ii) The band-alignment of Ge on Si is of type-II, *i.e.* only holes are confined inside the nanostructure, while electrons are weakly localized in the Si crystal around the QDs, leading to reduced matrix elements for radiative transitions.
iii) The lattice mismatch between Ge and Si is relatively small (about 4.2%), and thus, the Ge/Si QD dimensions tend to be too large for efficient zero-dimensional (0D) quantum confinement.

Consequently, light-emission from Ge/Si nanostructures is typically weak and restricted to cryogenic sample temperatures only.

In this chapter, we discuss a path for efficient room-temperature (RT) light-emission from Si that is based on QD material for which the optical properties are enhanced by ion implantation [19-22]. While such emitters have highly intriguing but to-date insufficiently-well understood structural, electronic and optical properties, we argue that this approach has the intrinsic potential to be a solution to the missing bottleneck in Si photonics: It solves the above-addressed deficiencies of group-IV nanostructure light-emitters and enables efficient light-emission even up to RT and above, still keeping the beneficial group-IV material compatibility to integrated Si technology.

## 1.1. Background

The elemental semiconductor Si and its oxide $SiO_2$ are by far the most important materials in integrated electronics as all integrated circuits are based on them. The ability to down-size active devices such as CMOS-transistors led to a massive increase in computer performance over the last decades and has been thus the main driving force for the worldwide digital revolution. However, the ever-increasing amount of global data traffic has led to two fundamental problems: Bandwidth limitations and escalating power consumption.

A higher packing density of CMOS-transistors requires a steadily increasing density of copper-based electrical interconnects. The small cross-section of down to ~100×100 nm$^2$ of the kilometres-long copper interconnects that carry the electronic signals in a single chip leads to substantial power dissipation due to Joule heating as well as parasitic capacitance and wiring delay, limiting the data transfer rates and bandwidth of integrated circuits. By using light instead of electrical current for data transfer, Si photonics can become a disruptive technology that aims at avoiding an explosion in energy consumption and at boosting data transfer rates at all levels of the interconnect hierarchy, from rack-to-rack to inter-chip and even intra-chip by creating low-cost, high-bandwidth optical interconnects.

Building a Si photonics platform or photonic integrated circuits (PIC) that are compatible with Si microelectronic technology requires several basic optical components such as waveguides [23-25], switches [26], filters [27], multiplexers [28], modulators [29,30], and detectors [31,32]. The development of these silicon-on-insulator (SOI)-based components are now, after more than a decade of extensive research, advancing rapidly and can be produced by Si-compatible technologies. However, the arguably most fundamental building block for such a universal platform remains elusive, due to several shortcomings of Si as a photonic material and mainly due to silicon's indirect bandgap: A practical, cost-efficient, and electrically pumped group-IV laser source for monolithic integration on a Si chip and that can be easily coupled to standard Si waveguide technology and operated at room temperature (RT) and above has yet to be demonstrated. This ultimate bottleneck for Si photonics was already anticipated by Soref and Lorenzo in the mid-eighties as they wrote:

*"From the foregoing information, we infer that single-crystal Si will be suitable for building directional couplers, filters, star couplers, optical switches, mode converters, polarizers, interferometers, and modulators that operate at λ = 1.3 or 1.6 μm (and beyond)-essentially every integrated-optical component except an optical source."* [33]

Further benefits of efficient light generation from Si would be that sensors, using Si-based near-infrared and especially mid-infrared (NIR & MIR) light-emitters merged with the power of Si microelectronics, could lead to many disruptive applications in *e.g.* environmental and atmospheric monitoring [16,17].

Approaches for overcoming the lack practical Si-based light emitters include bulk Si approaches [34], using dislocation loops in crystalline bulk Si [35,36], quantum confinement in group-IV nanocrystals [37-41], rare earth doping [42-45], Raman lasers [46,47], SiGe lasers by the engineering of Ge layers towards a direct bandgap by applying strain [48-51] or alloying with Sn using SiGeSn layers [10-13,52,53], hexagonal Ge and SiGe alloys [54,55] and last, but not least group-III-V lasers, bonded [56] or epitaxially-grown on Si substrates [14,15,57,58]. All of those approaches have both advantages and disadvantages, the latter mainly concerning the emission wavelength being above or below the telecommunication range, the temperature stability of the light-emission, especially in the temperature range from room temperature (RT) to 100°C, the possibility of electrical injection, parasitic recombination due to innovative technology concepts, continuous-wave-operation (cw), the threshold for lasing, state-of-the-art waveguide coupling and implementation into photonic circuits due to the use of extensively thick buffer layers, CMOS-process compatibility, requirements concerning substrate orientation and substrate size, device yield, power consumption, footprint and device lifetime. Probably the most important point concerns the overall costs for large-scale integration of the lasers. This is mirrored in the long-term success of integrated Si technology. For almost all microelectronic devices there exist counterparts, made from different materials than Si with higher single device performance. But there is a common saying that "*if it can be done with silicon, it will be done with silicon*", owing to the efficient large-scale integration possibilities of Si technology.

### 1.2. In a nutshell: Potential light sources for Si photonics

At present, semiconductor light-emitting materials that can be easily implemented on Si platforms are understandably gaining attention as a potential game-changer in the previously mentioned efforts towards on-chip and inter-chip data communication. From this perspective, the last two decades have seen a surge of reports on various material combinations and designs, pushing the limits of what had previously been considered impossible. In sections 1.3 and 1.4, we will briefly review the respective advantages and disadvantages of different approaches for Si-compatible light emission along with core challenges currently limiting their development and benefits of their practical deployment. As this is a highly active research field, red brick walls that were thought as being unsurmountable have been torn down within the last couple of years [59]. While, for the sake of conciseness, this certainly cannot be an exhaustive list, we believe that the brief summary will give the interested reader a valuable overview of the recent progress in the development of practical Si-based lasers and point out the ongoing need for a concerted effort to overcome present limitations.

### 1.3. All-group-IV approaches

→ **Bulk silicon:** At the turn of the last century, emission from bulk Si at RT was demonstrated by reducing the density of non-radiative recombination center, surface passivation, and advanced doping profiles [34] and in another approach by limiting the carrier diffusion length by introducing dislocation loops introduced upon ion implantation [35,36,60-63] or other defect centers in bulk silicon [64]. Besides emission yield considerations, the fundamental drawback of using bulk Si as the light-emitting material is the emission wavelength at ~1100 nm, causing problems regarding light absorption and losses in standard $Si/SiO_2$ waveguides.

→ **Porous Si** [37,65,66] and **Si nanocrystal** [38-41,67,68] approaches aim to use spatial charge carrier confinement to enhance the radiative emission from Si through quantum confinement and increased wavefunction overlap. While optical gain was reported [40], the main issues are based on the performance of electrically–driven devices and the light-emission energies well above the Si bandgap, the latter being problematic for light-coupling and propagation in Si/SiO$_2$ waveguides.

→ The challenges of **epitaxial SiGe nanostructures on Si** and methods to overcome their limitations regarding light-emission will be discussed later in sections 2 and 3.

→ In **rare-earth doping** approaches, rare earth atoms such as erbium are introduced as optical recombination centers into Si, SiO$_2$ or Si nanocrystals [42-45]. While the emission wavelength of 1.55 µm is naturally perfectly matched to the telecommunication C-band, long radiative lifetimes in the ms range [69], limited electrical carrier injection due to the surrounding dielectric matrix, and high pumping thresholds make on-chip electrically-driven lasing challenging [43].

→ **Stimulated Raman scattering** in photonic crystal resonators (PhC) was used to demonstrate optically-pumped lasing with low thresholds and small footprints [46,47,70]. However, Raman-scattering-based lasers are intrinsically not ideal for on-chip applications as Raman scattering is always connected to optical pumping.

→ **SiGe bandgap engineering** aims at lasing from group-IV structures. By applying tensile strain to the material, the small energy separation between the indirect and direct bandgap of bulk Ge can be reduced, or the band-ordering even reversed [48-51,71-73]. Recently, lasing from tensile-strained micro-patterned Ge was reported [50] up to a sample temperature of about 100 K. Additionally, heavy n-type doping of Ge [74] can be used to induce electrically-pumped lasing under pulsed excitation from bulk Ge structures [49]. In any case, the indirect transition remains close to the direct one and is, thus, always a concern of possible losses, especially the mandatory operation temperatures for devices. Additionally, if the applied strain is too large, the emission wavelength shifts above the target wavelength of 1550 nm.

→ By alloying **Ge with Sn** [10-13,52,53], another group-IV material, an optically-pumped laser was first demonstrated under pulsed excitation [10]. This group-IV laser marks a breakthrough because of its direct bandgap, but still has severe drawbacks concerning on-chip integration. As in the case of the Ge lasers above, the indirect bandgap is always in the energetic vicinity of the direct one. Nonetheless, recently optically-pumped lasing up to a sample temperature of 270 K [75] and electrically-pumped lasing up to 100 K were reported [76]. However, in all cases, strong emission quenching sets in already at cryogenic conditions, likely due to the energetically close indirect bandgap. The laser operation wavelength is generally larger than 2 µm [10,12,53,75,76] which makes this approach rather suitable for sensing in the mid-infrared wavelength region than for on-chip applications.

→ The formation of a direct energy bandgap in group-IV materials has been predicted and was indeed observed in **hexagonal Ge and SiGe alloys** [54]. This approach makes use of Brillouin zone folding [77] by alternating two types of atomic stacking in Ge and in Si-Ge alloys in the form of a hexagonal crystal lattice. This can be achieved by the overgrowth of a group-III-V seed crystal that has the form of hexagonal GaAs wires, grown on GaAs (111)B substrates [55]. For those wires, the high surface-to-volume ratio enables the formation of metastable crystalline phases such as hexagonal Si or Ge. At present, the light-emission energies can be tuned from 0.35 eV to ~0.67 eV, *i.e.* well above the standard telecommunication wavelengths, making this approach probably more relevant for applications in chemical sensing by tracing the footprints of molecular vibration modes in this energy range. Additionally, arrays of such direct bandgap- nanowires could be in principle envisioned as potential material for mid-infrared detectors e.g. in lidar applications [54].

### 1.4. Group-III-V on group-IV approaches

A straightforward corollary to avoid the intrinsic indirect bandgap nature of the group-IV elements Si, Ge and their alloys is to implement inherently bright light-emitters such as many of the compound semiconductors of the group-III-V or group-II-VI that exhibit a direct bandgap. Some of these hybrid group-III-V on Si prototype lasers represent the state-of-the-art concerning their performance characteristics. In a way, also the above mentioned hexagonal SiGe approach can belong to this section since group-III-V nanowires are needed to obtain optically active hexagonal group-IV material [54,78].

→ **Bonding III-V lasers onto Si substrates** by either wafer or die-bonding aims at combining the optical properties of group-III-V elements with the strength of Si electronics [79,80]. Wafer bonding is fast but has several disadvantages: GaAs and InP wafers are typically of 100-150 mm in diameter, while standard Si and SOI wafers are 300 mm and, in future, possibly even 450 mm in diameter. This mismatch leads to a striking loss in material yield. Additionally, this approach is cost extensive as group-III-V wafers are more expensive than Si and the bonding of whole wafers is an inefficient use of the group-III-V material.

→ **Die-bonding** with die sizes >1mm$^2$ circumvents some of those disadvantages. However, also the dicing step can result in small damage at the edges responsible for imperfect bonding. Additionally, this process is cost-intensive because it is more time-consuming than wafer bonding. Coupling to Si/SiO$_2$ waveguides is not straightforward but can be done if the optical mode is shared between the gain region and the waveguide region [56].

→ **Transfer printing** of group-III-V materials on Si or SOI is more efficient than wafer bonding in terms of group-III-V material usage and it provides higher throughput and certain parallel fabrication possibilities as compared to die-bonding [81-83]. Impressively, photonic integrated circuits consisting of group-III-V material on Si has been demonstrated [83]. So far, limiting factors are the group-III-V coupon size that can be handled (<100 µm) while for larger coupon sizes detrimental stress can be induced in the active devices due to the necessary underetching of the structures [82].

→ **Monolithic integration of group-III-V semiconductors** on Si benefits from the excellent properties of the group-III-V constituents but faces issues with respect to the material implementation. Most prominent are uncontrolled and detrimental crystal defects originating from different crystal structure and strain at the Si/Ge and group-III-V interface. Differences in thermal expansion coefficients of the group-III-V and group-IV constituents lead to additional defects once the substrate is cooled down to RT. Notably, all these defects can potentially contribute to device lifetime limitations. The yields of monolithically grown devices do not yet reach those of heterogeneous ones and for all following approaches compromises and disadvantages have to be accepted. In general, the major drawback of implementing group-III-V materials on a Si platform is that they do not offer the benefits and cost advantages inherent in monolithic Si integration and CMOS technology.

→ One of the most promising group-III-V on Si approaches uses **InAs/GaAs QDs** as laser gain material [14,15,57,58,84,85] for lasing at an emission wavelength of ~1.3 µm. The QDs are grown on several micrometer-thick group-III-V buffer layers to reduce the density of threading dislocations. However, the thick buffer layers impose a severe drawback concerning the optical coupling of the group-III-V device to the rest of the photonic integrated circuit, *e.g.* Si/SiO$_2$ waveguides. Additionally, dislocations can be present even in the QD-layer region leading to non-radiative recombination of minority carriers and gain compression effects as observed in directly modulated Si-based QD lasers [86].

→ **III-V nanowire lasers** grown on Si [87-90] show promise as the strain can be effectively accommodated due to the small footprint of the wire. Thus, group-III-V nanowires can be grown virtually defect-free on silicon [91] and electrical pumping of nanowire lasers in the visible wavelength range was reported in 2014 [92]. However, for obtaining vertical nanowires, (111)-type substrates are needed which are not employed in CMOS-technology and, thus, routes towards efficient large scale integration possibilities still have to be pursued.

→ For **InP based devices** [93,94] RT lasing was demonstrated under pulsed optical excitation. Here, the group-III-V material is grown in defined nanotrenches for defect trapping. While this approach does not need micrometer-thick buffer layers and has been demonstrated to be compatible with 300 mm Si wafers [95], there exist some drawbacks. In general, the emission wavelength of 880 nm - 1040 nm is not in the telecom regime (leading to waveguide losses) which can be overcome by incorporating InGaAs QDs on the InP buffer layers, as reported recently [96]. Continuous-wave emission and electrical injection remain critical issues. While the latter was demonstrated for InGaAs on InP on Si lasers recently [97], the technology is to-date performance-wise less mature as compared to the aforementioned InGaAs on GaAs on Si lasers while in this case still suffering from the need of micrometers-thick group-III-V buffer layers [97].

→ For **GaSb/Si and GaN/Si material systems** lasing was demonstrated [92,98-101]. For the former, the main drawbacks concerning integrated photonics applications are high current densities, the long emission wavelength >2 µm and the need for micrometer-thick buffer layers [98,99], for the latter, the short emission wavelength of 370 nm - 450 nm [92,100,101].

→ **Template-assisted selective epitaxy** or TASE [102-105]. A flexible approach regarding device geometry that includes the incorporation of group-III-V bulk materials on Si was developed over the recent years by IBM [102-105]. Small seed openings are defined on the Si device layer of an SOI substrate which are directly overgrown by epitaxial group-III-V crystals. The small footprint of the seed-openings allows for defect confinement at the interface. Thus, TASE can avoid the need for thick SiGe or group-III-V virtual substrate to accommodate the accumulated strain caused by the epilayer-to-substrate lattice mismatch. The formation of $SiO_2$ shells of almost arbitrary shape in combination with selective group-III-V epitaxy by metal-organic chemical vapor deposition (MOCVD) allows for intrinsic adaptability regarding device shapes, *e.g.* for co-planar integration of group-III-V on Si devices [104], mandatory for the formation of photonic integrated circuits. Lasing action under picosecond-pulsed optical pumping was observed for GaAs microdisks on Si fabricated using TASE, exhibiting low thresholds for lasing of about 14 pJ/pulse [105]. Electrically injected lasing has yet to be demonstrated and is currently likely limited by the formation of defects at the original substrate interface. Further, bulk gain material has severe disadvantages as compared to QD lasers, especially concerning temperature stability and lasing threshold current [106]. It remains to be seen if group-III-V QDs with high optical quality can be formed by TASE for future Si photonics-based quantum dot lasers.

While the Si industry has been opened up regarding the diversity of elements of the periodic table from about 5% before the nineties (H, B, O, Al, Si and P) to about 45% nowadays, an aloofness and inertia towards severe changes comes naturally for such a large sector of industry (market share >400 billion US dollars [107]), as material innovations always significantly contribute to the complexity and thus to the costs. Thus, it remains to be seen if one of the before-described strategies to implement foreign III-V laser materials with Si microelectronics will be able to enter the mass market. This short review of the state-of-the-art makes it very clear that the search for new, disruptive approaches to create electrically pumped group-IV light-sources, operating above RT for practical monolithic integration on a Si platform needs to be continued.

## 2. Epitaxial group-IV nanostructures on silicon

In the early nineties, defect-free group-IV epitaxial quantum dots embedded in crystalline Si were first discovered by Mo et al [108] and Eaglesham and Cerullo [109]. This sparked great hopes in the scientific community that quantum effects in low dimensional nanostructures can be the game changer to modify the inherently poor light-emitters Si and Ge into decent ones so that they can be utilized for applications compatible with Si integration technology [24]. The description – defect-free – was seen of particular importance since many kinds of defects, such as misfit dislocations, stacking faults etc. can be electrically active, and, thus, leading to detrimental non-radiative recombination in low dimensional group-IV material [110,111]. Until today, the most research was devoted to the epitaxial growth of thin Ge layers on the technologically relevant Si(001) surface, although also other configurations such as Si epitaxy on Ge substrates [112,113,114] and SiGe and Ge on SOI [115,116,117] and Ge on Si(111) surfaces [118,119] were heavily investigated. It can be foreseen that in the near future significantly more research will be devoted to investigating also epitaxial Sn-based group-IV nanostructures on Ge and Si [120] due to the impact of GeSn bulk-based light-emission [10] and the aforementioned, group-III-V laser community-related considerations regarding the advantages of QD-lasers over to bulk lasers [106].

The formation of epitaxial Ge/Si QDs follows the Stranski-Krastanow growth mode [121] and can be in short explained as follows: When thin films of Ge (crystal lattice constant 5.65 Å) are deposited on a Si(001) substrate surface (lattice constant ~5.43 Å), the first few monolayers of Ge form a pseudomorphically-strained (~4.2%) two-dimensional wetting layer (WL), for which partial strain relief is provided to the formation of surface reconstruction [122,123], *i.e.* rebonding events of Ge atoms at the surface layer. As this provides only a limited amount of strain relieve, the total strain energy is increasing almost linearly with the increasing film thickness. At a certain critical thickness (~3-5 monolayer (ML), depending on the Ge growth temperature), it is energetically favorable to expose a larger amount of surface atoms by the formation of three-dimensional objects – quantum dots – benefitting thereby from a significant lowering of the strain energy in the QDs, see Fig. 1.

Enormous efforts have been undertaken to optimize the fabrication schemes for defect-free epitaxial nanostructure growth, *i.e.* strain-relieve occurring only elastically, as opposed to plastically by the introduction of dislocations. There along, the thin-film deposition of Ge on Si (along with the growth of InAs on GaAs) has become the model system for strain-driven nanostructure formation in the Stranski-Krastanow growth system [124,125]. This naïve picture would suggest that a suitable deposition of Ge on a Si surface leads to a formation of a nanometer-sized QD of pure Ge surrounded by Si. But it became clear very early that this description is by far too simple. Along the way, a myriad of growth-related phenomena were discovered which found their counterparts in other strain-driven thin film growth systems [125] such as some group-II–VI systems like e.g. CdSe/ZnSe [126], nitride systems such as InN/GaN [127], metal systems like Au/Ni, or Au/Ag [128], to ice on platinum [129].

The Ge on Si(001) heteroepitaxial growth system is particularly rich when it comes to architectures of the QDs and islands. The different shapes are determined by low-index, low energy crystal facets such as the {105}, {113}, {15 3 23}, {111} or {12 5 3}-facets [108,130-132]. Depending on subtle variations in the growth parameters such as deposition temperature and -rate, the composition and deposited amount of material, different shapes were found to appear predominantly, such as unfaceted mounds [133], pyramids [108,134], multi-facetted domes [130,135], or structures with larger height to base aspect ratio such as barns [131] and cupolas [132], continuously approaching the shape of a half-sphere.

The initial formation of the nanostructures occurs on a supersaturated wetting layer [135-137], *i.e.* when pure Ge is grown on Si(001) by MBE at 700°C and a low growth rate of 0.05 Å/s, the WL thickness initially exceeds its equilibrium thickness of roughly 3.2 ML by more than 30% (~1 ML) [135-137]. Then, counterintuitively, the first QD forming are not of the smaller, more shallow pyramids, but larger, but energetically more stable dome-like QDs with base diameters around 100 nm and heights of 20 nm [135]. Likely owed to the fact that the SiGe/Si is a binary system, as compared to ternary systems such as InGaAs/GaAs, a particularly rich theoretical and experimental insight was gained with respect to the physical aspects of QD formation phenomena. This includes the impact of two important structural parameters that crucially determine the electrical and optical properties of any heteroepitaxial system for which a lattice mismatch exists: The strain and the spatial chemical composition. Obviously, these two are linked to each other. As the QDs evolve, areas of Si and Ge start to mix through an efficient alloying mechanism that is triggered by surface diffusion. Such intermixing often leads to a Ge-lean QD base and a Ge rich apex [138-140] or outer shell [141,142]. The study of the chemical composition of individual SiGe QDs with sizes of only tens of nanometers in length and height was developed through two rivalling, yet complementary experimental methods: X-ray diffraction based on synchrotron radiation with sub-micrometre resolution [138] and nanotomography, a method of alternating SiGe composition sensitive selective chemical etching and atomic force microscopy [139]. Being much more than a mere note of spectacular material science, these local composition gradients strongly influence the localization and overlap of wavefunctions [140,141] and thus directly influence recombination paths in the QDs [141] as Ge-rich inclusions at the nanostructure apex can lead to the formation of small and strongly quantum confined QDs inside larger nanostructures [140,143].

Furthermore, overgrowth of QDs with Si can influence the QD shape, internal SiGe composition [137,144,145] and thus the optical properties [137,146,147]. While overgrowth with Si at low temperatures (<~500°C) preserves the QD properties to a large extent [137,145], capping at higher temperatures (>~500°C) can lead to flattening of larger QDs and even to a shape transformation of steeper domes to more shallow pyramids [137,148]. Peculiar phenomena such as Ostwald ripening between QDs, coarsening, and shape transformations by growing and shrinking nucleation cores [149] have been reported. Many more phenomena related to the growth of Ge and SiGe QDs on Si have been observed and can be found *e.g.* in the following reviews and the references therein. [6,123,125,150]

As the structural parameters, also the optical and electrical parameters of Ge/Si QDs have been a subject to intense investigations [7,8,18,24,110,135,137,140,151]. The spatial location of the electron- and hole wavefunctions are indicated in the sketch in Figure 1(a). In photoluminescence (PL), electron-hole pairs are excited in the crystalline Si matrix surrounding the QDs. Holes relax to the energetic minima in the QDs, where they are strongly confined by the large valence band (VB) offsets of ~700 meV between the Ge QD and the surrounding Si matrix. Thus, depending on the size and chemical composition of the QDs, the quantum confinement energy experienced by holes and the energy of the VB offsets can vary. For many Ge/Si QD structures, the thermal escape of holes from the VB potential to the surrounding Si is be associated with high activation energies >300 meV. Thus, ionization of the QD through hole-escape is hardly possible at temperatures that are relevant for many optoelectronic device applications, *i.e.* from about 0°C to 100°C. Therefore, for the VB alone, the electronic properties in Ge/Si QDs are very promising concerning light-emission and similar to those of the best semiconductor QD light-emitters such as InGaAs/GaAs QDs.

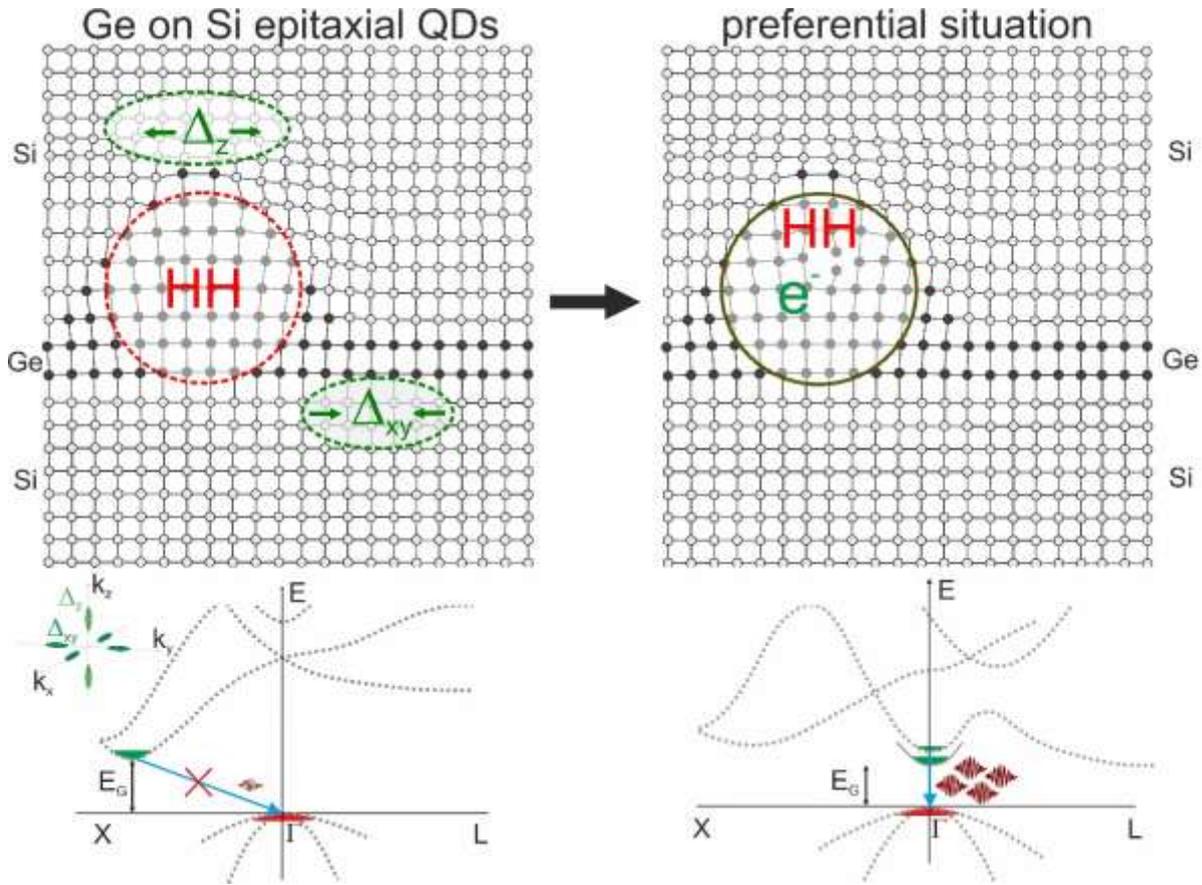

*Figure 1 Up: Schematic representations of strain-relaxed Ge QDs (black) in a Si matrix (grey). Upper left: A defect-free QD. The respective locations of the ground states for the electrons ($\Delta_z$ and $\Delta_{xy}$) and heavy hole states (HH) are indicated. Upper right: Preferential electronic configuration, achievable through the selective introduction of crystal defects. Bottom: Schematic band structure indicating optical transition paths (light blue arrows). Bottom, inset: splitting of the 6-fold degeneracy of the electronic minima in the CB upon applied strain. Bottom left: Band structure indicating the weak radiative transition from Si-like electronic states at the $\Delta$-points to Ge-like hole states at the $\Gamma$-point for conventional QDs. Bottom right: Preferential optical recombination at the $\Gamma$-point. Radiative enhancement can be expected due to direct bandgap transitions and the large spatial overlap of the wavefunctions (type-I charge carrier confinement).*

The real problem of these group-IV light-emitters originates from the properties of the conduction band (CB). The global energetic minima for the electrons are quantum confined states in approximately triangular potentials. These are introduced by strain in the Si matrix that is induced by the elastic relaxation of the Ge QDs [152-154]. Thus, electron wavefunctions barely enter the QD and this spatial separation [154], a type-II band alignment, leads to a reduced wavefunction overlap between the electron- and the hole states. To make matters worse, these strain-induced potentials in the Si matrix are energetically very shallow. Depending on the QD size, an activation energy of only up to about 100 meV is obtained for the thermal escape of the electrons [140,154,155] which necessarily leads to pronounced PL quenching, already at cryogenic sample temperatures. Additionally, both, Si and Ge are indirect semiconductors and optical transitions have to be assisted by phonons or scattering events. Thus, albeit all the efforts in understanding the physical nature of GeSi/Si QDs, the overall PL efficiency is rather poor, especially as compared to the best direct bandgap group-III-V nanostructures. Thus, over the past fifteen years, strategies have been employed to improve the overall light emission-yield from Ge/Si QDs. This includes the fabrication of site-controlled QDs with well-defined inter-QD-distances [7,18,110,140,147,156,158] or by coupling the light-emission to photonic resonances of micro-resonators and plasmonic structures [159-165].

Significant improvement concerning light-emission was reported for randomly nucleated Ge/Si(001) QDs for which a subsequent photonic resonator was created using standard lithography and etching techniques. These resonator designs include e.g. photonic crystals (159,161,162,164) or microdisks [19,160] and microrings [160]. An improvement of the optical properties can also be based on plasmonic effects, by coupling the QD layer beneath the Si cap to a plasmonic nanoparticle that is deposited on top of the Si capping layer [166,167]. Naturally, for these resonator structures, no particular alignment to the QD positions is possible using self-assembled nanostructures. The resonance behavior in photonic microcavities is characterized by their cavity modes for which the location varies within the cavity. This, consequentially, leads to problems regarding the coupling between the cavity modes and randomly nucleated QDs for which no, or only minimal control over the microscopic, spatial nucleation position can be obtained. Therefore, after subsequent resonator formation, the QD positions and the positions of highest field enhancement will only coincidentally overlap and it is not possible to guarantee ideal QD cavity-mode overlap for ten thousands of optical devices that should be produced in photonic integrated circuits.

Group-IV nanostructures, and in particular, Ge/Si QDs, however, offer a valuable solution to this problem, namely deterministic site-control of QDs [7,9,110,147,156,157]. Different approaches for defining QD nucleation sites have been investigated in the past two decades. These include e.g. selective epitaxy using $SiO_2$ masks [11,117,168], scanning tunneling nanolithography [169], and local implantation by focused ion beams [170,171]. Many works reported on the use of etched pits for defining preferential nucleation sites, during both molecular beam epitaxy (MBE) and chemical vapor deposition (CVD) growth [7,18,110,140,147,156,157,158]. It was demonstrated that QD nucleation into periodic pit arrays can result in an improved morphological and chemical uniformity [140,141,147,172-174], which is associated with substantially narrower ensemble photoluminescence (PL) emission line width [7,110,140]. Single SiGe QDs that are grown on pit-patterned Si substrates with wide inter-QD distances were used to study the optical response of individual QDs in this group-IV system [156,175]. At the lowest excitation powers employed (100 nW), a full-width-at-half-maximum linewidth of only about 16 meV of the QD-related no-phonon peak emitted at 920 meV was reported, which is a record low for such epitaxial Ge/Si QDs [175].

Combining site-control of QDs and resonator formation allows for a deterministic placement of the QDs with respect to the resonator modes. Such a high degree of controllability was recently demonstrated by placing Ge/Si QDs at various position of L3 cavities of a photonic crystal resonator to increase the light emission from defined resonator modes [164]. Additionally, in this way an experimental mapping of the photon density of states of a resonator can be performed, allowing for a direct feedback loop between theoretical simulations and experimental realization of the devices [164].

In summary, the understanding of structural and optical properties of Ge/Si QDs and the coupling to microcavities brought significant progress in the field of CMOS compatible light-emission using the flexible approach of Ge/Si heteroepitaxy. However, these methods of nano-optical light enhancement through photonic resonances seem to be a mere workaround that cannot answer the fundamental question: How to obtain light emission from epitaxial group-IV quantum dots that can be comparable to that of group-III-V direct bandgap material, at least in the temperature window between 300 K and 370 K. These QDs would be, through their structural properties, environmentally benign and, at the same time, chemically robust, perfect candidates for active material in optoelectronic devices due to their fabrication flexibility and intrinsic compatibility with Si microelectronic standards. In other words, what is really needed for epitaxial Ge/Si nanostructures, is a paradigm change.

In the following, we will address a promising way to improve the intrinsic optical properties of Ge nanoclusters upon implantation of heavy ions. Ion implantation is one of the indispensable processes in microchip fabrication, as only implantation of donors such as phosphorus and acceptors such as boron allow us to control the electrical properties of Si and to make devices such as transistors. Almost all semiconductor manufacturing processes today make use of at least two but often many more ion implantation steps, *e.g.* for almost all doping in silicon integrated circuits. Most commonly, B, As, P, Sb, In, Si, Ge N, He and H are implanted. Ion implantation offers high controllability and precision in incorporating the dopants at the designed location within the semiconductor device. For this, the implantation depth crucially depends on factors such as the mass and energy of the dopant ion, the mass of the target atoms, the implantation angle and the orientation of the target as well as the implantation temperature [176].

Concerning optical emission from Si, the implantation of ions, such as H, *etc* into crystalline bulk Si or other bulk group-IV layers was studied quite extensively within the last decades and led to the observation of interesting implantation-related light emission phenomena from these structures [35,36, 177-180]. Further research was devoted to nanostructure formation driven by ion implantation [181-184]. Little, however, was reported about the direct implantation of ions into already existing, small, nanometer-sized nanostructures that are embedded in a crystalline host matrix. This is surprising, since both, nanostructure formation and ion implantation have shown to provide benefits concerning optical and optoelectronic properties as compared to bulk Si. Following, we will discuss that what is needed to obtain efficient light from Si is interlacing these two approaches.

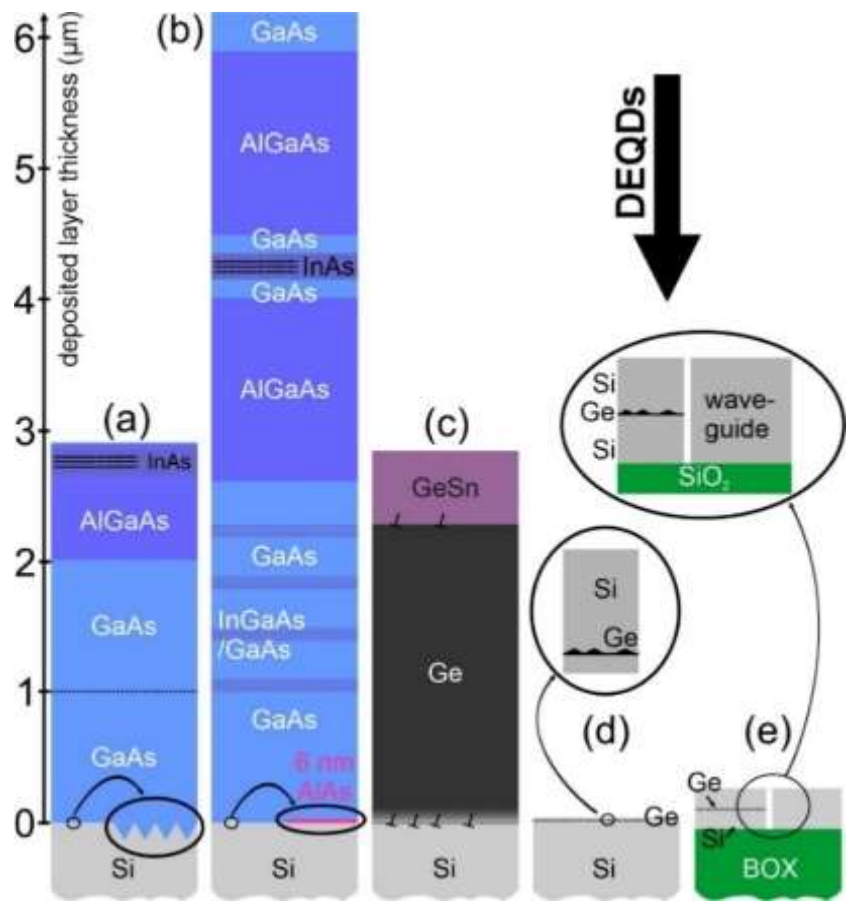

*Figure 2 A comparison of possible monolithic implementations of semiconductor light sources on Si substrates. (a) and (b) Growth of multi-stacked InAs QDs on GaAs grown directly on Si. [15,58] (c) Growth of bulk GeSn or GeSiSn heterostructures on top of a Ge virtual substrate grown on Si [10]. (d) Defect-enhanced GeSi quantum dots (DEQDs) grown directly on Si substrate [19,20] or (e) on silicon-on-insulator (SOI) substrates. A scheme for evanescent coupling of DEQD light-emission to SOI waveguides is indicated in (e). Image modified from [18] with permission from the authors.*

## 3. Ion implantation into Ge quantum dots on silicon

We will focus on the investigation and fostering of a novel material class, ion-implanted Ge/Si QDs that eventually might be the game changer in providing a cost-efficient, practical, and electrically-driven laser, emitting in the telecom wavelength range at ~1.3 and 1.55 µm, and operating at RT and above. The main point favoring such a technology would be that these lasers can be compatible with the standard SOI photonics platforms, *e.g.* the SOI 220 nm device layer platform. Due to the absence of thick buffer layers, no complicated coupling to waveguides needs to be employed (see Fig. 2), in contrast to other material combinations (see sections 1.3 and 1.4). The need for thick buffer layers for the GeSn or group-III-V materials on Si (see Figs. 2(a)-(c)) is one of the bottlenecks for these approaches. While for the latter, the light sources themselves have excellent output characteristics, the simple and low-loss coupling of the light source to Si photonics waveguides remains an open issue. A scheme of the different layer stacks employed for the different material approaches is shown in Fig. 2. From this scheme, the flexibility and large-scale integration possibilities, as well as the benefit of employing Ge/Si-based light-emitters for on-chip or intra-chip data transmission, is evident as coupling to other photonic components such as photonic waveguides, filters, splitters, modulators, multipliers and detectors is rather straightforward (Fig. 2(e)).

Paradoxically, and as mentioned above, up to now, one of the main criteria for the fabrication of Ge/Si QDs and nanostructures, in general, was to control and avoid defect formation. For course, this is particularly true that some defects, such as 60° misfit dislocations in QDs. It was shown that the presence of these can fully suppress radiative recombination paths in the QDs [110]. But, as for bulk material, not all defects are detrimental for optical radiative recombination [35,36]. The main idea behind the additional ion-implantation into the nanostructures is to overcome the aforementioned shortcomings of conventional Ge/Si QDs (Fig. 1) by capturing holes and electrons in two fundamentally different but interlaced low dimensional structures: epitaxial QDs and extended point defects. Such unique QD modification can result in exquisite and, in particular, superior structural and optical properties. Thus, these nanostructures are referred to as *defect-enhanced Ge quantum dots* or DEQDs.

First, we introduce the most striking benefits of DEQDs, before elaborating on the later in section 3. Already at cryogenic temperatures, *i.e.* T <~ 200 K, the luminescence yield of DEQDs is vastly enhanced as compared to conventional, defect-free Ge/Si QDs [19], see Fig. 3(a). At RT, conventional Ge/Si QDs are in essence not emitting any light, while DEQDs exhibit almost the same light-emission yield as at 10 K. For DEQDs, also clear signs for optically pumped lasing are present [19], *e.g.* linewidth narrowing, threshold behavior and mode competition (Fig. 3(b) and Ref. [19]). The radiative carrier lifetimes in DEQDs can be very short, down to the range of less than 1 ns [19]. It is relatively straightforward to electrically contact DEQD layers. Consequently, DEQD light-emitting diodes (LEDs) that emit up to device temperature above 100°C [21] have been reported, which is exceptional for group-IV light-emitters. Due to the small volume of the active DEQD nanostructure material, it is possible to enhance the active gain volume by vertical stacking of the DEQDs [22], therefore increasing the light-emission yield in a scalable way related to the DEQD density [21,22]. Furthermore, we will discuss the surprising thermal stability of the structural properties of DEQDs which withstand thermal annealing at 600°C for at least 2 hours, or flash-lamp annealing at 800°C for at least 20 ms [185,186].

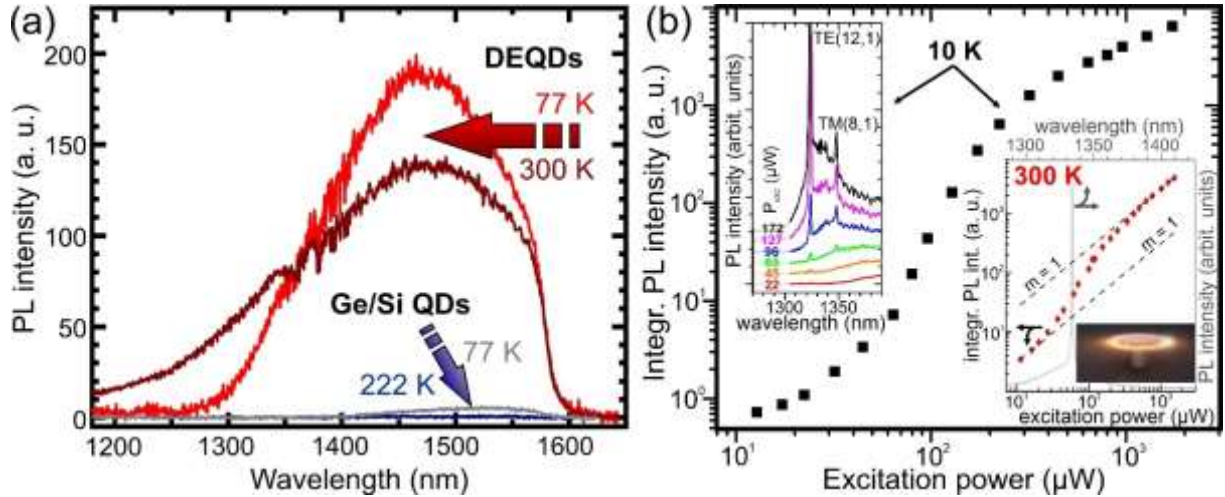

*Figure 3* *(a) Increase of the photoluminescence (PL) yield at 77 K and 300 K of DEQDs as compared to conventional Ge/Si QDs, grown under the same growth conditions but without additional ion implantation. For those, the PL fully quenches already at ~200 K. (b) Signs for optically-driven lasing at 1330 nm from microdisk cavities containing DEQD-gain material. The integrated PL intensity versus excitation power is depicted on a double logarithmic scale and exhibits a clear s-shape and indicates lasing at 10 K (black squares) and 300 K (red-diamonds, right inset). Left inset in (b): Emergence of whispering gallery modes upon increased optical pumping. A schematic image of the underetched microdisk is shown in the bottom-right inset in (b) Linewidth-narrowing and mode competition from these resonator modes are discussed in Ref. [19]. (b) Image modified from [19] with permission from the authors.*

### 3.1. DEQD fabrication procedure

The current understanding of the DEQD formation process is depicted in Fig. 4. Si(001) or SOI (001) substrates are *ex-situ* chemically cleaned before their transfer to a solid-source molecular beam epitaxy (MBE) chamber. *In-situ*, the samples are degassed, usually at temperatures of about 700°C for ~15 min. Hereafter, a Si buffer layer of about 50 to 100 nm thickness is deposited to ensure a clean, epitaxially-grown sample surface before the deposition of a thin Ge film. About 4.5 to 7 monolayers (ML) of Ge (*i.e.* 6.3 Å to 9.9 Å) are deposited at typical growth rates of 0.03 Å/s to 0.1 Å/s and a growth temperatures $T_G$ typically between 400°C and 700°C. First, a two-dimensional wetting layer (WL) forms, followed by elastic relaxation of the strained layer that results in QD formation. Depending on the growth temperature, different QD shapes are obtained, as discussed above in section 2. In the following, we will mainly focus on Ge growth at $T_G = 500°C$ which results in the formation of elongated pyramids, also called huts or hut-clusters [108]. They depicted in Fig. 4 as light-grey shape. At any point during the growth, a negative bias voltage between 0 V and up to 3 kV can be applied to the substrate. Since Ge and Si layers are evaporated from electron beam evaporators, Ge and Si ions are always present in the chamber as a certain fraction of the evaporated atoms again passes through the electron beam, leading to atom ionization. These positively charged ions can then be intentionally implanted into a specific epilayer, by turning on the substrate bias at a certain time of the growth.

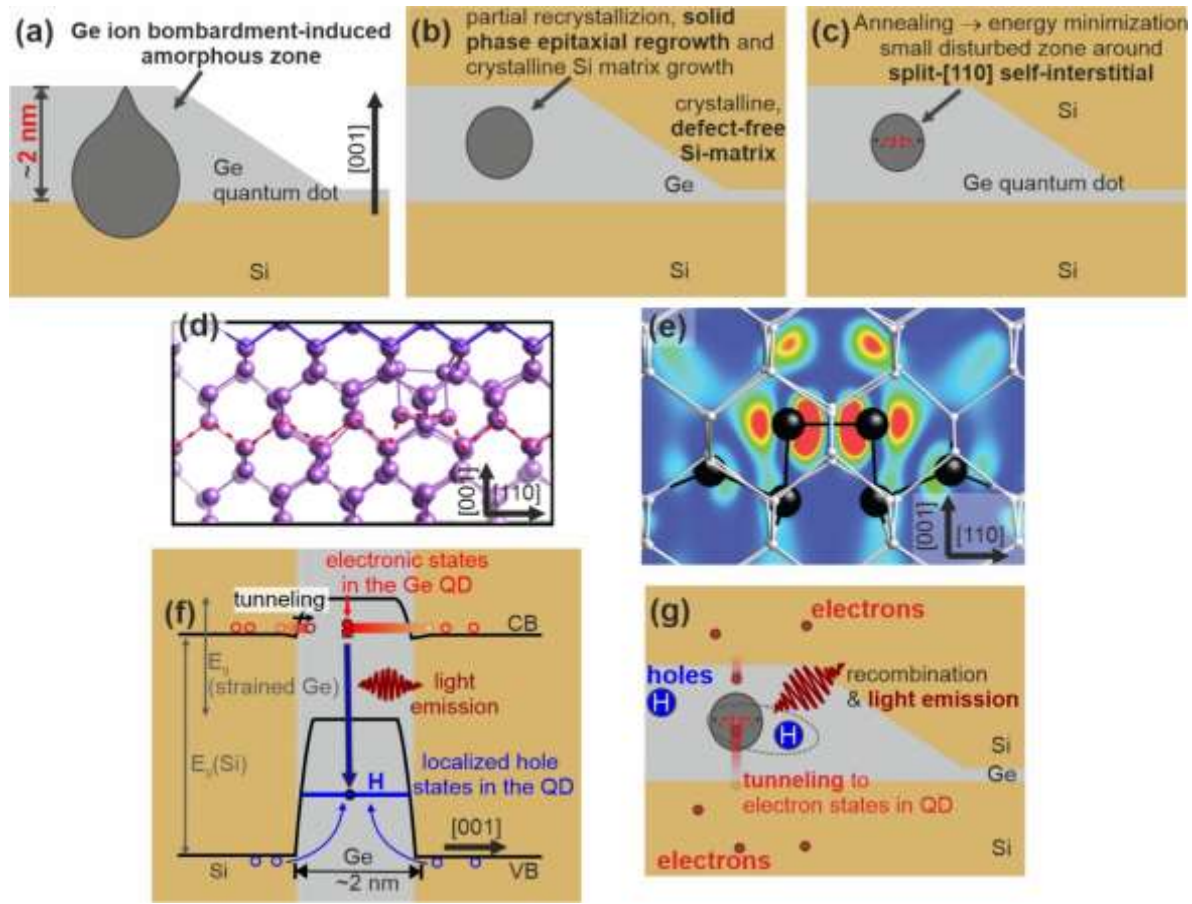

*Figure 4 (a)-(c) Split-interstitial formation in Ge QDs upon Ge ion implantation and annealing. (d) Ground-state defect structure that results after computational crystallization of an amorphous state containing one extra Ge atom. The blue-solid and red-dashed line highlight the undisturbed lattice and the area containing the split-[110] self-interstitial. (e) DEQD crystal lattice including the split-[110] self-interstitial (black). The color code presents the highest (red) and lowest (blue) electronic orbital electron density of 0.00035 and 0.0 electrons/bohr$^3$, respectively. This indicates a strong modification of the electron-states due to the defect-site. (g)-(h) Proposed light-emission processes in DEQDs (f) as an energy level scheme and (g) in real space. Electrons tunnel to the defect-induced states in the QD, recombining with holes confined in the QD leading to direct recombination in real and reciprocal space. Image modified from [20] with permission from the authors.*

In order to create DEQDs, the substrate bias is either applied during the whole Ge deposition or during the later stages of Ge QD growth, so that the low-energy Ge ions (<3 keV) impinge on the QD with a dose that was estimated to be ~$10^4$ µm$^{-2}$. Thereby, the energy of the Ge ions is significantly higher than the displacement energy of Ge atoms in the crystal that is about 14 eV [187,188]. Thus, even the first displaced atoms, so-called recoil atoms, have sufficient energy to displace further atoms and the resulting collision cascade is only stopped when the initial ion energy has dissipated [176]. Therefore, within these ~2 nm high hut clusters the heavy Ge ions with energies lower then 3 keV are causing shallow, about 1-2 nm large amorphized zones in the QD's crystal lattice, as displayed as dark-grey regions in Fig. 4(a). During the ongoing Ge growth, the Si capping layer deposition and possible post-growth annealing, solid-phase epitaxial regrowth (SPER) partially recrystallizes the amorphized zone [189]. Note, that perfect recrystallization is impeded since an additional ion was brought into an already perfect crystal lattice, see Fig. 4(b). Nevertheless, the recrystallization at the growth-front allows for subsequent crystalline and defect-free overgrowth of the DEQDs with Si [19,20] which is of particular importance regarding charge carrier confinement and light-emission and to avoid unwanted non-radiated recombination through surface states.

After the growth of a Si capping layer and, if applied, thermal annealing, a crystal configuration with low formation energy emerges out of the amorphized region in the QD, Fig. 4(c). Theoretical calculations suggest that for a strain-relaxed Ge crystal that contains N+1 atoms in a volume for which N atoms would lead to a perfect crystal lattice, the minimum energy crystal configuration is in the form of a split-[110] Ge-Ge self-interstitial [20,190-192]. The presence of such relatively large implanted ions must lead additionally to a pronounced deformation of the surrounding crystal lattice, involving about 45 neighbouring atoms [20]. Figure 4(d) depicts the calculated ground-state defect structure of an amorphous state containing one extra Ge atom [20] that results after an extensive series of Monte-Carlo quench-and-anneal steps and subsequent geometry relaxation [20] using the Quantum Espresso DFT package [193]. The blue-solid and red-dashed line in Fig. 4(d) highlight the calculated undisturbed lattice and the area containing the resulting split-[110] self-interstitial. Figure 4(e) presents the core of the DEQD crystal lattice including the split-[110] self-interstitial (black) and the surrounding distorted crystal lattice (white). Overlaid in Fig. 4(e) is the electronic orbital isosurface cross-section indicating that the electron wave-functions are influenced by the presence of the defect-site (red color) due to electron-states at the Γ-point ~70 meV below the Ge CB edge [20]. The wave-functions of the holes seems to remain largely unaffected by the crystal distortion [20] and more details on the defect-induced band structure changes, especially band structure effects in the whole Brillouin zone and the influence of strain will have to be evaluated in the future [194].

**3.2. Light-emission from DEQDs**

Our present understanding of the light-emission dynamics from DEQDs is depicted in Figs. 4(f) and (g) as an energy band diagram (Fig. 4(f)) and in real space (Fig. 4(g)), respectively. Under non-resonant optical- or electrical excitation, free carriers are generated mostly in the Si matrix barriers surrounding the DEQDs from where they thermalize to the band edge minima of the Si conduction band (CB) and valence band (VB). Holes, indicated in blue in Figs. 4(f) and (g), are efficiently trapped and confined in the DEQD VB potential as in conventional Ge/Si QDs. Note, that the average hut-cluster height is only about 2 nm, due to the very shallow {105} facets of the QDs.

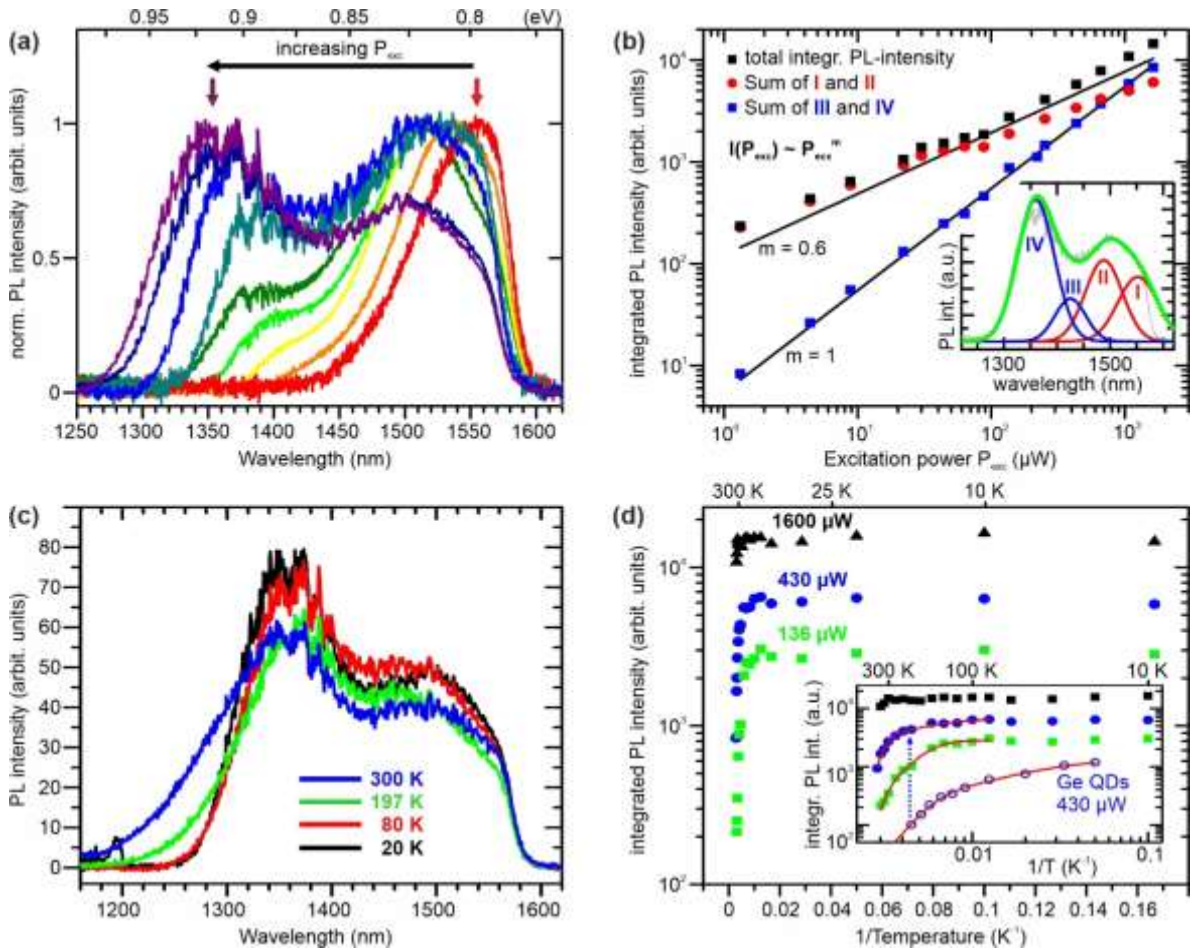

*Figure 5 (a) Normalized PL spectra recorded at RT and for increasing laser excitation power ($P_{exc}$). A pronounced blue-shift with increasing $P_{exc}$ is observed (b) The integrated RT-PL intensity $I_{PL}$ vs $P_{exc}$ is plotted for the higher-energy peak (blue squares) the lower-energy peak (red spheres) and the total DEQD emission (black squares). The black solid lines represent power coefficients m = 0.6 and m =1, according to $I_{PL} \sim P_{exc}^m$. (c) PL spectra of DEQDs for $T_{PL}$ of 20, 80, 197, and 300 K. The spectra are not normalized. (d) Full symbols: $I_{PL}$ of the DEQDs vs inverse $T_{PL}$ for $P_{exc}$ = 136, 430, and 1600 µW. The red curves are Arrhenius fits to the data. The open blue circles show $I_{PL}$ of Ge-QDs without ion-implantation treatment vs $1/T_{PL}$ for $P_{exc}$ = 430 µW. Image modified from [19] with permission from the authors.*

Thus, in the worst case, the split-[110] self-interstitial is located in the vertical center of the QD, *i.e.* at most, 1 nm away from the Si matrix layer. For such short distances, electrons, indicated in red in Figs. 4(f) and (g), can tunnel from the Si layers through the ~1 nm thick QD-induced barrier to the QD with its defect-site from where they can radiatively recombine with holes [20]. Since electrons and holes are confined within the DEQD and both carrier types can occupy states at the Γ-point [20], and the radiative transitions are direct in real and reciprocal space, in stark contrast to conventional QDs.

From the experimental evidence, we expect that defect-induced changes will lead to increased light-matter coupling, but the matrix elements for dipole-allowed transitions $\propto \langle f | \hat{e} \cdot \vec{p} | i \rangle$ (with $|i\rangle$ and $|f\rangle$ the initial and final electronic state, $\vec{p}$ the electron momentum, and $\hat{e}$ the light-polarization vector) were not reported so far [194]. In the following, we will discuss the experimentally observed optical properties of DEQDs that are in agreement with the above-described light-emission scheme.

Figures 5(a) and (b) depicts the influence of increasing PL excitation power ($P_{exc}$) on the light emission from DEQDs. First, a pronounced blue-shift of the peak-PL emission from about 0.8 eV (1550 nm) to 0.925 eV (1350 nm) is observed for increasing $P_{exc}$. Interestingly, there exists an inner structure within the PL signal that can be separated into a lower energy and a higher energy signal, as can be seen in Fig. 5(b). In a previous publication, we tentatively assigned the blue-shift to filling of hole states with increasing $P_{exc}$ in the ensemble of measured QDs [19]. In Figure 5(b) the integrated RT PL intensities $I_{PL}$ of the total DEQD emission as well of the higher energy and lower energy peak are plotted versus $P_{exc}$. The lower energy peak increases with a power factor m of about 0.6 according to $I_{PL} = P^m$. This power factor is commonly associated with Auger-recombination, and often observed in conventional GeSi QDs [141,195,196]. Noteworthy, the higher energy peak rises with $P_{exc}$ according to a power factor m of 1, which is assigned to optically direct recombination in semiconductors [197].

The probably most spectacular feature of the optical properties of DEQDs is depicted in Fig. 5(c). The PL spectra were recorded at different sample temperatures $T_{PL}$ of 20, 80, 197, and 300 K. Note that these spectra are not normalized which immediately implies that temperature-induced PL quenching is basically absent in DEQD samples, at least in a temperature window from 4 K to RT and at elevated $P_{exc}$ [19]. To shed more light on this behavior, in Fig. 5(d) the detailed temperature-dependence of the DEQD PL signal is plotted in an Arrhenius-type graph with the PL intensity quenching at high $T_{PL}$ being fitted using Arrhenius fits, see solid red lines in Fig. 5(d) [19]. In the inset of Fig. 5(d) the data are plotted on a double logarithmic scale to emphasize the PL quenching behavior at high $T_{PL}$ that is associated with activation energies $E_A$ of ~350 meV for both $P_{exc}$ = 136 and 430 μW. This is in stark contrast to what was observed in conventional Ge QDs for which thermal quenching with $E_A \sim 60 - 80$ meV were reported [24,198] (see inset of Figure 5(d)). As discussed above, for conventional Ge/Si QDs, electrons condense at the global energy minima of the strain pockets in the Si matrix that are induced by the relaxation of the QD. These potentials are energetically shallow and, thus, already at cryogenic temperatures, light-emission quenching occurs, associated with the thermal escape of electrons to the surrounding Si.

In contrast, our present data suggest that thermal quenching in DEQD light-emission only sets in when holes can thermally escape from the QD valence band potential to the surrounding wetting layer (Ref. [186] and see Section 3.6). This process is associated with an activation energy of 300–400 meV [19,186], making DEQDs suitable light-emitters even above RT. Additionally, the tunnelling process for electrons from the Si matrix to the QDs is temperature–independent and electron-hole recombination times have been reported to be very short [19]. Thus, the already long lifetime of electrons in the surrounding Si matrix [24] is further enlarged if holes are efficiently collected by the DEQD potentials. This could prevent detrimental electron-hole recombination in Si and ensure that electrons are always available to tunnel to the DEQD from where light-emission in the telecommunication regime occurs upon radiative carrier recombination. These processes make the DEQD system highly suitable even for device temperatures even up to 100°C, as will be seen in the following section 3.4.

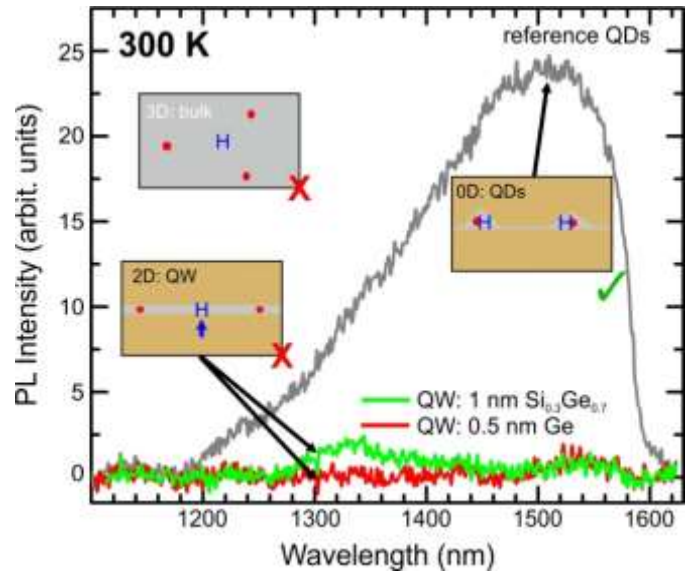

*Figure 6* *Room-temperature PL spectra of DEQDs (gray spectrum) and two different quantum well samples (QWs) (green and red spectra) for which Ge ion implantation was performed under the same conditions as for the DEQDs. Only the DEQDs show strong PL enhancement. Inserts: Schematic views of DEQDs (right) versus ion-implanted QW (lower left) and a Ge bulk sample containing split-[110] interstitial defects. On the one hand, holes (blue H) are confined in DEQDs and thus overlap strongly with electrons trapped at defect-sites (red dots). On the other hand, holes in an ion-implanted QW or in bulk Ge can diffuse away from the two adjacent point defects (blue arrow), leading to a reduced the electron-hole wavefunction overlap. Image modified from [20] with permission from the authors.*

The carrier recombination scheme presented in Figs. 4(f) and (g) is consistent with the observed DEQD emission wavelength from 1300 nm to 1600 nm (Fig. 5), the high activation energies for thermal quenching (Fig. 5(d)), a power-coefficient of m=1 (Fig. 5(b)), the negligible thermal PL-quenching of the PL-intensity at RT (Figs. 5(c,d))and the dramatically shortened carrier lifetimes down to ~0.6 ns [19]. Our present understanding is that excitation of an ensemble of DEQDs with different sizes as well as filling of different defect-induced energy levels leads to the observed broad spectral emission range. Furthermore, in Ref. [19] it was found that the energetic difference between the calculated ground states and excited states for heavy holes in an ~2 nm high hut cluster are about 100 meV apart which would also agree with the energetic splitting between the high-energy and low-energy peak, observed for DEQD samples. Further research will be necessary to unambiguously track down the origin of multi-peak light-emission from DEQDs under high PL pumping power.

For DEQDs, a natural question concerns the role of the surrounding environment of the defects – the QD cages that host the defects. Can this approach would also work for other structure with different dimensions? In a recent publication [20], we found that superior light-emission is only possible if the defects are introduced into QDs, as indicated in Fig. 6. There, the room-temperature PL response of a DEQD sample (grey spectrum) is compared to two different quantum wells samples (QWs) (green and red spectra) for which Ge ion implantation was performed under the same experimental conditions as for the DEQDs. Pronounced light-emission is only observed for the DEQD sample while in the ion-implanted QW samples the optical properties at RT are dramatically reduced. We understand that when QDs, with their finite spatial boundaries in all three dimensions, are the host-surrounding of the defects, then holes, indicated in blue color in the scheme in Fig. 6, are confined within the QDs because the QD potential imposes the global energy minimum in the valence band. Therefore, the hole wavefunctions are forced to overlap strongly with the electron wavefunctions that are attracted by the defect-induced changes in the conduction band at the QDs site (red dots), leading to direct light-emission from DEQDs. In contrast, in quantum wells and Ge bulk samples containing the same defects, we argue that even though the defect clearly seems to be attractive concerning electron recombination, the additional presence of strain around the defect-site respells the holes from the QD-site. In this way, holes can move to more strain-relaxed regions in-between two such defects [20]. The resulting charge carrier separation reduces the spatial overlap of the electron-hole wave-function and thus the transition matrix elements for optical transitions. Hence, enhanced PL yields at RT can neither be observed in control samples in which Ge-ions were implanted into QW structures (Fig. 6), nor in bulk Ge [20], where split-[110] self-interstitials are well-known for decades [190-192]. Thus, quantum dots are essential for DEQD light-emission and only the incorporation of the defect into the QD with nanometer precision in growth direction makes the superior optical properties of DEQDs feasible. Therefore, for preferential defect formation, the defect must be created in the QD and, thus, implantation depth, and thus the ion energy, ion species and QD height have to be considered simultaneously [176].

### 3.3. Considerations towards large scale integration possibilities:

Up to now, the DEQDs were grown by molecular beam epitaxy (MBE) combined with *in-situ* Ge ion implantation. For any industrial applications, where large-scale integration plays a role, a transfer of the growth of the avtive DEQD material to CMOS-compatible fabrication techniques has to be considered.

This can be generally envisioned as Ge/Si QD growth can be achieved in high structural and optical quality also by chemical vapor deposition, CVD [2,3]. Furthermore, ion implantation and rapid thermal annealing are standard processes in Si technology. To ensure defect formation inside the QD, implantation has to probably occur on uncapped samples or samples with a few nanometer-thick Si capping layer. Otherwise, it might be technologically challenging to ensure implantation of the ions into the QDs through several 100 nm thick layers due to the resulting smeared out implantation profiles. A transfer of the DEQD schemes to CVD methods seems to be feasible but still requires the necessary adaptation of machinery for CVD growth and in-situ transfer to the low energy ion implantation to avoid detrimental contamination. Modern industrial implanters face still challenges concerning the productivity of the beam current at energies below 10 keV. But since device scaling demands for ion implantation at such low energies, this is likely to be solved by industry. Also, the implanter-induced cross-contamination has to be monitored and controlled as this can significantly influence the DEQD defect formation. Concerning the thermal budget, CMOS-compatibility for backend implementation is guaranteed by the low DEQD growth temperatures of 400-500°C [18-22].

### 3.4. Electrical injection:

Even though RT electroluminescence has been reported from conventional Ge/Si QDs [24,199,200], the reported light emission has been subject to extensive thermal quenching and, thus, their potential as practical RT light-emitters is limited by all the aforementioned reasons. For DEQDs, the only significant change in the fabrication procedure concerns the additional ion implantation step into the QD layer. As conventional Ge/SiQDs, DEQDs are surrounded by a crystalline and nominally defect-free Si matrix, which can be easily doped p-type or n-type using *e.g.* boron or phosphorous or antimony. Thus, the DEQD-system is ideally suited for electrical injection *e.g.* by growing them into the intrinsic region of a p-i-n diode [21]. For such LED devices, the most essential observation is the outstanding temperature-stability of the optical properties of DEQDs even under electrical pumping conditions. For the first generation of DEQD-LEDs, vertically stacked layers of QDs were employed that were placed in the middle of a 200 nm thick intrinsic region of a p-i-n silicon diode [21]. The p-type and n-type regions were doped with $2.5 \cdot 10^{19}$ cm$^{-3}$ of boron and about $1 \cdot 10^{19}$ cm$^{-3}$ of arsenic, respectively. To determine the scalability of the DEQD-density on the LEDs, multi-DEQD layers of three and seven stacks were fabricated together with an empty Si diode, *i.e.* without DEQDs. The device geometry of the 100×100 µm$^2$ sized mesa structure is depicted in the inset of Fig. 7(a) [21], where also to top Al metallization layer and the bonding wires are indicated. In Figure 7(a) the electrical characteristics of an LED with seven-layers of DEQDs is presented for an injection current density of 10 kAcm$^{-2}$ and recorded at heat sink temperatures around 300 K and above [21]. No drastic change in the spectral emission characteristics was observed in this range of sample temperatures. Note the robust LED operation at heat sink temperatures as high as 100°C (373 K) for a duty-cycle up to 5%, which remarkably for group-IV light emitters. In the upper inset of Fig. 7(a), the light-current characteristics are presented for the heat sink temperature range from 22°C to 100°C.

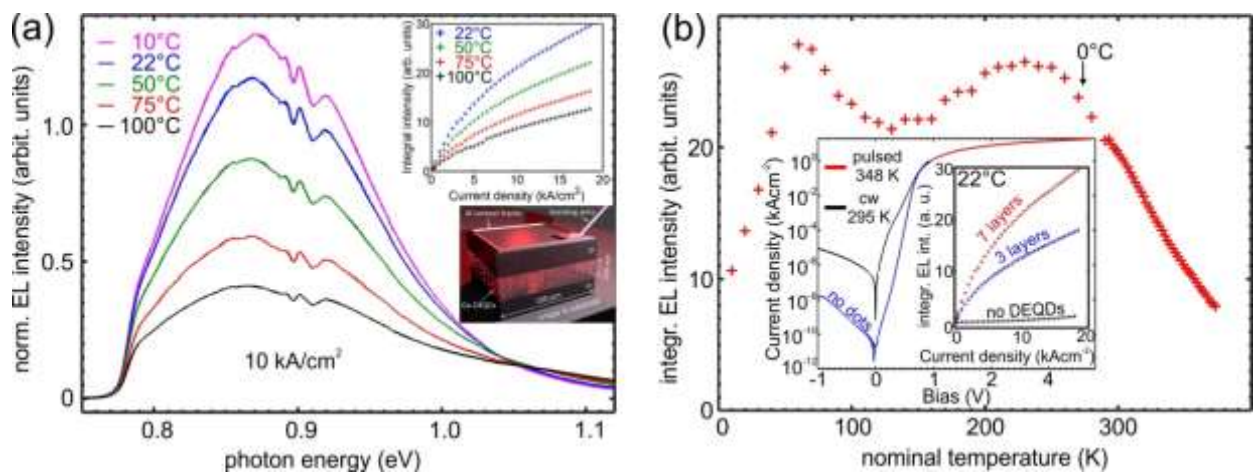

*Figure 7* *(a) Spectrally-resolved evolution of the electroluminescence (EL) emission properties of an LED containing seven DEQD layers around and above RT in Peltier-controlled operation. The upper inset depicts the light-current characteristics for high heat-sink temperatures. The lower inset presents a scheme of the device geometry. (b) The integral intensity of the DEQD-LED (driving current density 10 kA/cm$^2$) vs. the heat-sink temperature. Emission quenching sets in above 230 K, but the LED performs well up to at least 100°C (373 K). The first inset shows the T-dependence and I-V-characteristics of the DEQD-LEDs using continuous-wave (cw) (black and blue curves) and pulsed excitation (red curve). The second inset shows the low-current behavior of the DEQD LEDs (blue and red) and the Si reference (black), respectively, measured under cw operation. Good scaling of the integrated emission intensity with the number of incorporated DEQD layers is demonstrated. Reprint (adapted) with the permission from P. Rauter et al. ACS Photonics 5, 431 (2018). Copyright (2018) American Chemical Society.*

Noteworthy, the maximum values for the heat-sink temperature and driving current were in Ref. [21] only limited by either by the Peltier temperature controller or by the strength and number of bonding wires and the pulsed current source. No significant change in the shape of the curves with increasing current and only very weak light-emission saturation is observed, implying that DEQDs allow for stable light emission output under strong electrical pumping (at least 20 kAcm$^{-2}$) and very high sample temperatures.

In Figure 7(b) the integrated electroluminescence signal from the 7-layer DEQD-LED is plotted versus the heat sink temperature. The intensity remains at about 75% of the global maximum in a cryogenic temperature range from 60 K to 230 K before an actual emission quenching sets in that is associated with the thermal escape of holes, as discussed earlier in section 3.2. As a comparison, for optically-pumped DEQD structures (Fig. 5), emission quenching at high pumping powers is delayed to sample temperatures higher than 300 K. The earlier onset of emission quenching from LEDs is likely related to the higher actual device temperature at a specific heat sink temperature due to ohmic heating induced by the driving current. Nevertheless, the integrated intensity still remains at about 30% of the global maximum for an impressive heat-sink temperature of 100°C (373 K), which is extremely late as compared to what is reported for other direct bandgap systems such as tensile-strained Ge of GeSn or SiGeSn materials [10,12,13,50]. The superior temperature stability of the DEQD light-emission is of particular importance for electrically-pumped laser resonators. These have to be fabricated on buried oxide layers to ensure optical confinement of the laser modes with respect to the gain material. In this case, the buried oxide permits efficient heat transfer between the laser and the heat sink, emphasizing the urgent need to employ temperature-stable gain material, such as DEQDs.

The first inset in Fig. 7(b) compares the current-voltage characteristics of a 7-layer DEQD LED under low currents to the Si reference diode. Evidently, the reverse current is raised by about three orders of magnitude by the incorporation of Ge material in the form of DEQDs. However, in the relevant driving regime, *i.e.* under forward bias and current densities >100 Acm$^{-2}$ the presence of DEQDs hardly influence the diode characteristics. Joule heating obviously plays a role when comparing the current-voltage characteristics obtained using cw excitation to pulsed operation at high driving currents and it can be seen that heat sink temperature of 75°C under pulsed excitation corresponds to a heat sink temperature of 22°C under cw operation [21].

In the second inset of Fig 7(b), the light-current characteristics are plotted for the three-layer- and seven-layer DEQD-LED, as well as for the Si reference diode. The electroluminescence emission intensity is continuously increasing with the driving current up to the experimental limit at about 20 kAcm$^{-2}$, without any strong saturation effects being observed. Thus, it can be expected that higher driving currents will not impose a limiting factor for the future realization of an electrically-pumped DEQD laser. The Si reference diode only shows very weak emission, originating from the optical interband transitions in bulk Si. Increasing the amount of DEQD multilayers from three to seven, *i.e.* an increase in DEQD sheet-density by a factor of 2.3, increases the light-emission by a factor of 1.65 without changes in the shape of the light-current curve. The resulting emission power scaling by DEQD layer stacking in LED might turn out essential for the future development of electrically-pumped laser sources with DEQD-gain material. The slight deviation from a linear scaling behavior between light-emission and DEQD density is likely caused by charge carrier competition between the different DEQD layers, which can be mitigated by stronger electrical pumping of the devices and future optimization of the diode structure itself. No particular effort was invested by now to optimize doping concentrations and thickness of the contact and intrinsic regions to ensure ideal recombination currents in the DEQDs. Thus, improvements in this respect will be crucial for the fabrication of future DEQD-laser devices.

In summary, electrical pumping of DEQD emitters is clearly feasible and excellent optical properties of these unique group-IV dots are maintained under electrical pumping. Comparing the PL and electroluminescence date from DEQD samples shows that for the same collection efficiency and under maximum available pumping conditions the same time-averaged light-emission was obtained [21]. Considering the differences in the duty cycle between the results of Fig. 5 and Fig. 7, a twenty-fold higher emission intensity from the DEQD-LEDs was obtained as compared to the DEQD-emitters that show clear signs for optically pumped lasing (Figure 2(b)) if they are embedded in microdisk resonators [19,21]. Therefore, electrically-pumped lasing from DEQD gain material in ridge- or microdisk-resonators is expected to be possible, even at application-relevant temperatures.

### 3.5. Scalability of DEQD densities

In the previous paragraph, we have seen that vertical DEQD-stacking allows for very good scalability of the gain material volume which can be a significant factor concerning future laser applications. However, it has to be noted that vertical stacking of DEQDs is intrinsically challenging due to the necessary ion implantation in each Ge layer. As noted above, the low-energy ion implantation during the DEQD process results in an amorphous zone in the QD (Fig. 8(c)) that partially recrystallizes via solid-phase epitaxial regrowth during the growth interrupts and the continued deposition of Si and Ge (Fig. 8(d)).

For stacked DEQD layers, the vertical distance between the QDs is preferably small so that high gain material volumes can be achieved at a certain layer thickness, which can be obtained by restricting the Si spacer thickness to about 10 to 30 nm. However, if the recrystallization is not leading for every QD in a crystalline growth front, then during the following Si overgrowth defects such as stacking faults and polycrystalline regions will be continued in the Si spacer layer above the DEQDs. These defects allow for a relaxation of the compressive strain that is built up by the elastic relaxation of the former QD layers and thus, leads for the following Ge layer to preferential nucleation site, at which a majority of the deposited Ge accumulates (Fig. 8(f)).

This, again, leads to non-crystalline overgrowth with Si which further feeds the Ge material accumulation cascade in the next-next Ge layers and leads to a massive increase of the defective region (Figs. 8(g)-(h)). This process can be avoided by using elevated substrate temperature $T_G$ for the growth of the Si spacer layers. In Figure 8(a), the cross-sectional transmission electron microscopy (TEM) image of an 11-fold multi-stack of DEQD layer is depicted for which the spacer layer $T_G$s that were ramped up from 500°C to 600°C. On the contrary, if the spacer layer is grown at $T_G$ = 350°C to 500°C leads to the formation of massive defect structures in the later periods of the multi-stack as can be seen in the cross-sectional TEM image in Figure 8(b).

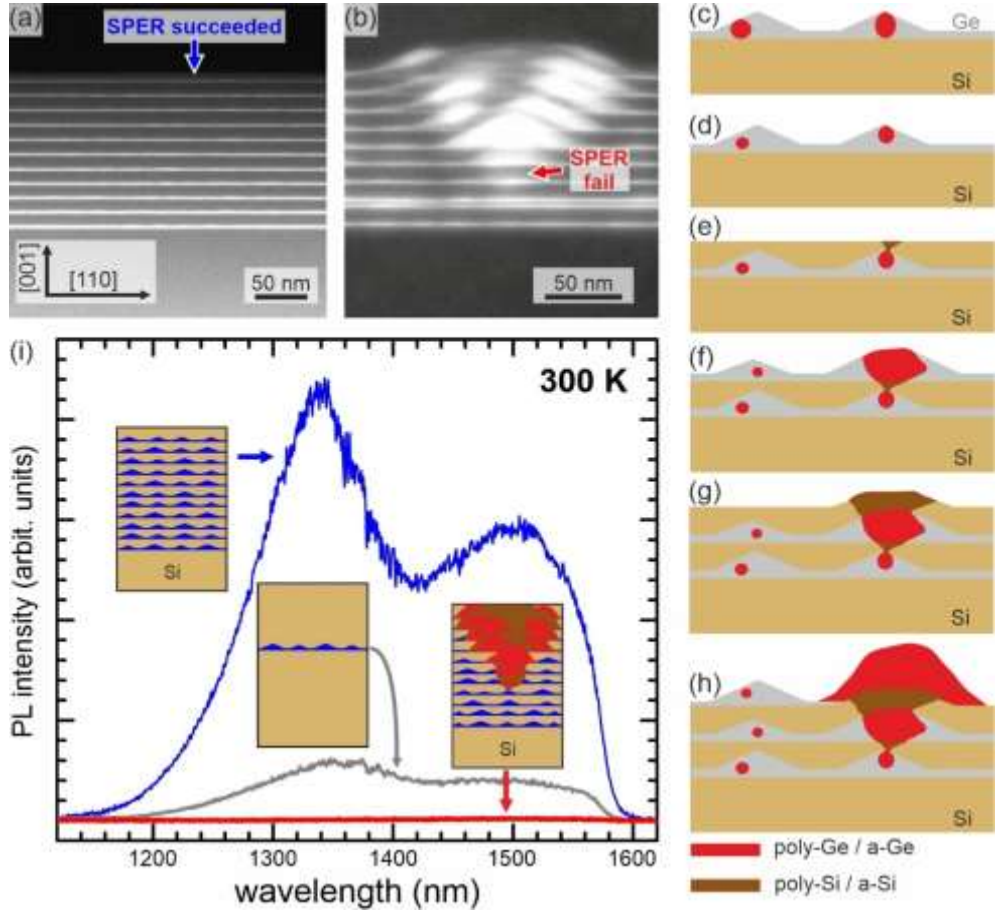

*Figure 8 (a) HAADF STEM images of an 11-fold stack of DEQDs (light grey) separated by 15 nm thick Si spacer layers grown at T = 500°C – 600°C. (b) HAADF-STEM image of a 10-fold multi-stack of DEQDs where the 12 nm thick Si spacer layers were deposited at T = 350°C – 500°C. (c)-(h) Schematic evolution of the massive defects formed upon SPER-fail. Non-perfect recrystallization of the growth front (d) leads to defect formation in the Si spacer layer (e) leading to Ge accumulation in the subsequent DEQD layers (f)-(h). (i) RT PL spectra obtained from a single layer of DEQDs (grey) and from an 11-fold stack (blue) for which T was 500°C – 600°C. For the multi-stack sample for which the growth parameters were not optimized (T = 350°C – 500°C), no PL signal was recorded (red). Image modified from [22] with permission from the authors.*

The need to avoid such accumulating defect structures is immediately evident by evaluating the PL response from the successful and failed layer configuration depicted in the cross-sectional STEM images in Figs. 8(a) and (b). Samples, where massive defects are present, do not show any PL signal at room temperature (red spectrum), even if the laser excitation and detection spots are several micrometers away from such a defect center. This implies that the density of states of non-radiative recombination centers must be very high in such defect clusters. Thus, the absence of DEQD-PL emission in the presence of the massive defects is not surprising considering that for Ge/Si QDs already single dislocations can fully quench the PL emission from nearby, but defect-free QDs [110].

On the other hand, when solid-phase epitaxial regrowth was successful for all layers of the multi-stack (Fig. 8(a)), then the light emission intensity can be increased by a factor of 7.5 as compared to a single layer of DEQDs, measured under the same experimental conditions, see and blue spectrum in Fig. 8(c). Likely, at the employed $P_{exc}$, DEQDs still compete for the generated electron-hole pairs leading to the observation that enhancement factor is slightly smaller than the factor by which the number of DEQDs increases for different samples. However, the total layer thickness between the single layer-sample and the multilayer sample in Fig. 8 remained the same in all cases. Thus, the growth of multilayer DEQDs allows for suitable scaling of the gain material volume, even for device applications, for which the target thickness is restricted *e.g.* to 220 nm as in standard photonics SOI technology.

### 3.6. Thermal budget and annealing of DEQDs

A natural question targets the robustness whenever the properties of low-dimensional structures crucially depend on the formation and existence of point-defects or defect complexes which consist only of a few atoms. For DEQDs, the question whether they are thermally robust enough to be employed in CMOS-compatible processes can be answered emphatically "yes", as it was shown that DEQDs retain their superior light emission properties, even under thermal annealing at 600°C for two hours [20,186] and 800°C for about 20 ms [185], which does not represent an upper boundary.

For the creation of CMOS-compatible photonic integrated circuits, monolithic light sources, such as DEQDs have the distinct advantage of possible front-end integration, *i.e.*, placing the photonics layer in-between the CMOS layer and the metallization layers. This integration possibility is favorable as compared to back-side or back-end integration since slow electrical driving of the integrated optics devices through vias (back-side) or the slowest metallization layers (back-end) can be avoided [35]. Of course, front-end integration requires that the thermal budget of the photonics layer must not deteriorate the underlying CMOS layer while being itself robust enough to withstand the fabrication of the subsequent metallization layers. The CMOS-layer can typically sustain thermal budgets of about 1 h at 450°C [201] or 0.5 h at 475°C, while AlCu, or Cu metallization layers are typically deposited in a temperature window of at least 350°C to 475°C [201,202].

Figure 9 depicts the spectral evolution of the DEQD low-temperature and RT-PL emission, following *in-situ* thermal annealing for 2 hours at temperatures $T_A$, ranging from 500°C to 675°C. For $T_{PL}$ = 10 K (Fig. 9(a)), a pronounced increase of the integrated PL emission is observed with increasing $T_A$, concomitant with a blue-shift of the peak-emission. Only if the samples underwent annealing at $T_A$ = 675°C, the PL emission abruptly quenches (Fig. 9(a)). A similar, yet qualitatively different behavior is observed it the sample temperature is raised to 300 K, Fig. 9(b). While the blue-shift with increasing $T_A$ is observed as well, the PL yield decreases already for $T_A$ > 600°C.

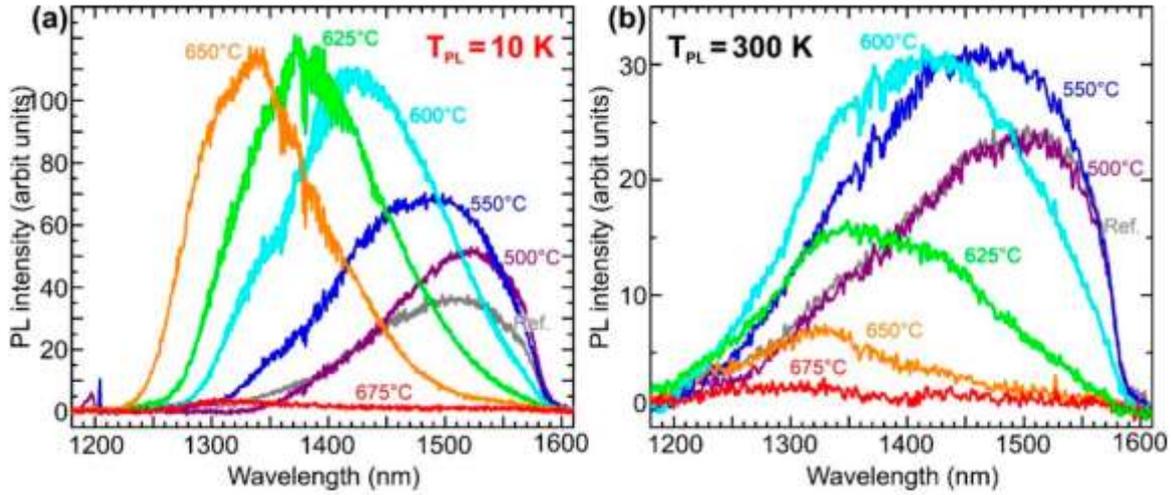

***Figure 9*** *PL spectra of DEQDs, recorded at (a) $T_{PL}$ = 10 K and (b) $T_{PL}$ = 300 K. The spectrum of the DEQD reference sample is shown in grey. The other DEQD samples were in-situ annealed for 2 h at annealing temperature $T_A$ of 500°C (violet spectra), 550°C (dark blue), 600°C (light blue) 625°C (green), 650°C (orange) and 675°C (red). Image modified from [186] with permission from the authors.*

From Fig. 9, two main conclusions can be drawn. First, the abrupt decline in integrated PL intensities after thermal annealing at high $T_A$ can be associated with the thermally-induced migration of the defect complex from the QD into the surrounding Si [20]. By evaluating an Arrhenius-fit of the integrated PL versus $T_A$ for $T_{PL}$ = 300 K, a lower boundary for the activation energy for defect migration $E_A$(migr) of 3.4 eV was found [186]. It was argued that the strong resilience of the thermally-induced migration of the defect complex is based on the stabilizing effects of the strained regions around the split-[110] self-interstitial core [20,186].

The second conclusion obtained from Figure 9 concerns the changes in the Ge composition profiles of the DEQDs with increasing $T_A$ and the associated reason for the thermal quenching of the PL at high $T_{PL}$. The spectral shift of the DEQD-PL emission with increasing $T_A$ (Fig. 9) in combination with the activation energy $E_A$ for thermal quenching with increasing $T_{PL}$ allows for modelling of the composition profiles of the DEQDs and the surrounding wetting layer as a function of $T_A$. The results, which include a thermal degradation of the Ge profiles of DEQDs and the WL through bulk diffusion are plotted in Figs. 10(a) and (b). The actual QD ionization path at high sample of device temperatures was found to be caused by the thermal escape of holes, confined in QD (Fig. 10(a)) to the energetically higher quantum-confined states in the surrounding wetting layer [186], Fig. 10(b). This escape process is schematically presented in the cross-sectional TEM image and as a top-view scheme in Figs. 10(c) and (d).

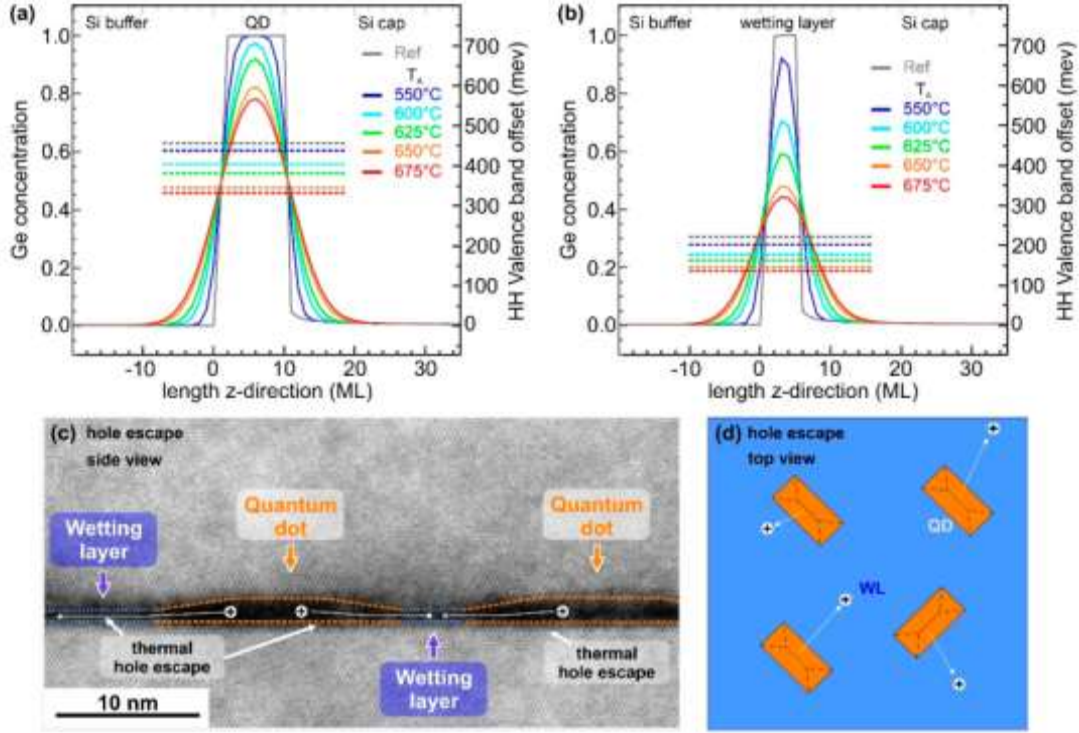

*Figure 10* Simulated Ge composition profiles, the respective valence band offsets, and the heavy-hole ground state energies (dashed lines) for different $T_A$ for (a) a QD of 1.4 nm height, and (b) the surrounding wetting layer (WL) (height 0.7 nm), respectively. (c) Side-view scheme and (d) top-view scheme of thermal hole-escape from the QD potentials to higher-energy states in the WL potential, leading to spatial electron-hole-pair separation and thermally induced luminescence quenching. Image modified from [186] with permission from the authors.

### 3.7. Curing and passivation of non-radiative recombination centers

In addition to the split-[110] self-interstitial, unwanted defects in DEQDs can arise from non-optimal sample cleaning before growth, growth of Ge and Si itself or the low-energy ion implantation conditions. For the latter, imperfect solid phase epitaxial regrowth (section 3.5) or an ion collision cascade that stops below the QDs can be detrimental to the optical properties since it might lead to the insertion of non-radiative recombination centers. These need to be avoided or, at least treated post-growth, so that their negative influence on the optical properties can be minimized. Furthermore, it was reported that the presence of extended crystal defects, such as misfit dislocations can be associated with PL quenching in group-IV emitters [110], making the interpretation of the observed changes of the PL yield with respect to the performed *in-situ/ex-situ* treatments impossible.

In the previous section 3.6, we have seen that post-growth thermal annealing can improve the optical quality of the DEQD-emitter by possibly recrystallizing such non-radiative recombination centers.

However, this can come at a prize of changed emission wavelengths (Fig. 9) originating from morphological and compositional changes due to the annealing (Fig. 10). As compared to in-situ thermal annealing or rapid thermal annealing, flashlamp annealing provides the advantage of high temperatures, *e.g.* for dopant activation or point-defect recrystallization while it avoids due to short exposure times severe structural changes that unavoidably also lead to changes in the optical properties, see Fig. 9 [186]. It was demonstrated that flash-lamp annealing at 800°C for up to 20 ms, performed on DEQDs leads to only negligible modifications of the electronic band alignment [185]. For such accurately chosen parameters for the flash-lamp treatment, the PL emission yield can be increased by almost 50% while not changing the spectral shape of the DEQD emission. What is more, it was shown that flash lamp annealing leads to improved stability of the PL emission intensity over a temperature range from 10 to 300 K [185]. Both, the increase in PL yield and the enhanced thermal stability is attributed to the thermal healing of unwanted and non-radiative defects in the Si matrix around the QDs. Overall, the results of sections 3.6. and 3.7. show that DEQDs can in principle be implemented into hybrid integrated circuits where high-temperature steps are necessary for the fabrication process.

In a different approach, hydrogen passivation can be employed to saturate the dangling bonds from unwanted defects in the Si matrix. This method is widely used in applications ranging from solar cells to CMOS technology to emission improvement of GeSi nanostructures [203,204,205]. In Figure 11, we compare the PL emission for a series of DEQD samples that underwent various low-energy proton irradiation treatment with different irradiation doses ($5\times10^{17}$ cm$^{-2}$, $1\times10^{18}$ cm$^{-2}$, and $5\times10^{18}$ cm$^{-2}$) to a DEQD reference sample (black spectra in Fig. 11). The spectra were recorded at sample temperatures of 300 K (Fig. 10(a)) and at $T_{PL}$ = 10 K for Fig. 11(b). The high-temperature spectra in Fig. 11 show that the DEQD emission yield can be increased by up to almost 90% for a low-energy proton irradiation dose of $1\times10^{18}$ cm$^{-2}$ (blue spectrum), as compared to the untreated reference sample (Figure 11, black curve). On the other hand, a 2.5-fold higher H-implantation dose (orange spectra in Fig. 11) leads to a 63% decrease of the integrated PL emission (Fig. 11(a)). The DEQD-PL emission spectra recorded at 10 K reveal (Fig. 11(b)), that the proton implantation causes additional carrier recombination paths that can be observed by the appearing emission band between 1250 nm and 1300 nm. This additional peak exhibits a systematic shift to higher energies with increasing proton irradiation dose. Due to its thermal deactivation behavior that is associated with a rather low activation energy of about 40 meV, it can be assigned to so-called G-centers that are ascribed in the literature to the formation of Si-C vacancy clusters in Si [206]. The presence of the G-centers suggests that structural changes are induced in the Si capping layer upon low-energy proton irradiation, whereas the PL-results confirm that the DEQD layer is structurally not affected by H-treatment [186].

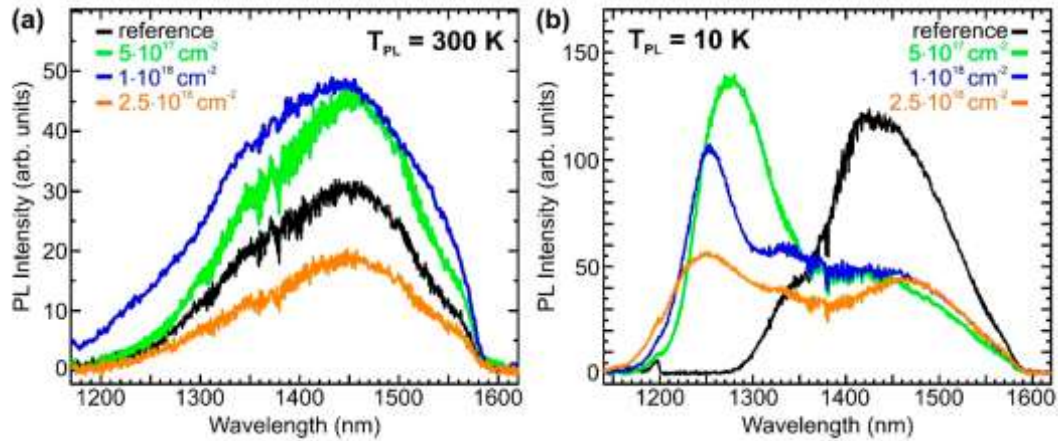

**Figure 11** PL spectra recorded at (a) $T_{PL}$ = 300 K and (b) $T_{PL}$ = 10 K of a reference sample annealed at $T_A$ = 600°C. In addition to the untreated sample (black curves), samples have been investigated that underwent H-implantation with implantation dose of $5\times10$ cm$^{-2}$ (green curves), $1\times10^{18}$ cm$^{-2}$ (blue curves), and $2.5\times10$ cm$^{-2}$ (orange curves). Image modified from [186] with permission from the authors.

Moreover, the light-emission improvement observed for the H-treated samples measured at 300 K proves that H-treatment is a valid tool to be used for group-IV nanostructures since non-radiative recombination in the Si matrix layers can be successfully suppressed. Further investigations should be targeted at employing H-treatment methods that omit proton implantation techniques, thus providing the benefits of the passivation of unwanted dangling bonds while not causing structural harm to any part of the device layer. As such, H-treatment techniques as used in the photovoltaic industry should be excellent candidates to achieve this goal [204,207]. There, H is supplied through the deposition of a H-rich SiN layer and then in a diffusion-driven process added to the Si crystal through advanced annealing techniques [207].

## 4. Summary and future directions in DEQD research:

Conventional wisdom, developed over decades of research on Si technology holds that microelectronic-grade quality requires (apart from doping) ultralow concentrations of impurities and crystalline defects that act as electronic traps. To this end, and in retrospect, the bright PL in the telecommunication range from QDs containing split-[110] self-interstitial defects could have been discovered much earlier; the formation of dislocation-free QDs was demonstrated about 30 years ago [108] while the split-[110] self-interstitial, an easily forming extended point defect, has been discovered also decades ago [190-192]. All that really needed to be done was to combine these two approaches. However, this idea would have been hard to conceive of since both constituents, the split-[110] self-interstitial in bulk and the defect-free epitaxial Ge/Si QDs show only weak or even no luminescence emission at room temperature.

In this respect, it seems to make quite some sense to revisit the combination of numerous group-IV nanostructure architectures and defect types, in terms of their light emission properties, both on theoretical as well as on experimental grounds. Future research will need to be devoted to obtaining electrically-driven-lasing, the microscopic understanding of the processes, leading to the enhanced light emission from DEQDs, including the influence of SiGe composition and strain on the optical properties.

For DEQDs, efficient light emission at room-temperature was demonstrated, both under optical and electrical excitation. The band structure modifications induced by the implanted defects lead to optical direct bandgap transitions and type-I band alignment, thus overcoming the main drawbacks of epitaxial Ge/Si QDs. When placed into photonic resonators, clear signs for optically pumped lasing using DEQD light emitters have been demonstrated. As a next step, the DEQD technology has to be transferred to an SOI platform, allowing also mode confinement in electrically-pumped devices. This would allow achieving electrically pumped lasing, ideally at room temperature, using DEQD gain material. Such laser sources will form the basis for a monolithic optoelectronic platform enabling superior data transfer rates and novel optical sensing functionalities for a vast number of applications. Despite the truly exciting feature of DEQDs, we believe that the preliminary research on these just scratched the surface of what is possible using DEQDs as Si light sources.

**Acknowledgements**

I would like to use this opportunity to gratefully thank the main mentors in my scientific career, Günther Bauer, Friedrich Schäffler, Thomas Fromherz and Armando Rastelli. This work was funded by the Austrian Science Fund (FWF): Y1238-N36, P29137-N36. Funding was also provided by the EU H2020 QuantERA ERA-NET via the Quantum Technologies project CUSPIDOR, which is co-funded by FWF(I3760) and the Linz Institute of Technology (LIT): LIT-2019-7-SEE-114.